\documentclass[pss,fleqn,embeddedheads]{w-art}
\usepackage{times}
\usepackage{w-thm}
\usepackage[]{graphicx,epsfig,amsmath}
\begin{document}
\DOIsuffix{theDOIsuffix}
\Volume{XX}
\Issue{1}
\Copyrightissue{01}
\Month{01}
\Year{2003}
\pagespan{1}{}
\Receiveddate{\sf zzz} \Reviseddate{\sf zzz} \Accepteddate{\sf
zzz} \Dateposted{\sf zzz}



\title[Shot Noise in Mesoscopic Transport]{Shot Noise in Mesoscopic Transport Through Localised States}


\author[A.K. Savchenko]{A.K. Savchenko\footnote{Corresponding
     author: e-mail: {\sf A.K.Savchenko@ex.ac.uk}, Phone: +44\,1392\,264109, Fax:
     +44\,1392\,264111}\inst{1}}
\author[S.S. Safonov]{S.S. Safonov\inst{1}}
\author[S.H. Roshko]{S.H. Roshko\inst{1}}
\address[\inst{1}]{School of Physics, University of Exeter, Stocker Road, Exeter, EX4
4QL, United Kingdom}

\author[D.A. Bagrets]{D.A. Bagrets\inst{2}}
\author[O.N. Jouravlev]{O.N. Jouravlev\inst{2}}
\author[Y.V. Nazarov]{Y.V. Nazarov\inst{2}}
\address[\inst{2}]{Department of Applied Physics, Delft University of Technology,
Lorentzweg 1, 2628 CJ Delft, The Netherlands}

\author[E.H. Linfield]{E.H. Linfield\inst{3}}
\author[D.A. Ritchie]{D.A. Ritchie\inst{3}}
\address[\inst{3}]{Cavendish Laboratory, University of Cambridge, Madingley Road, Cambridge, CB3 0HE, United Kingdom}
\begin{abstract}
We show that shot noise can be used for studies of hopping and resonant tunnelling between localised electron
states. In hopping via several states, shot noise is seen to be suppressed compared with its classical
Poisson value $S_I=2eI$ ($I$ is the average current) and the suppression depends on the distribution of the
barriers between the localised states. In resonant tunnelling through a single impurity an enhancement of
shot noise is observed. It has been established, both theoretically and experimentally, that a considerable
increase of noise occurs due to Coulomb interaction between two resonant tunnelling channels.
\end{abstract}

\maketitle                   




\renewcommand{\leftmark}
{A.K. Savchenko et al.: Shot Noise in Mesoscopic Transport Through Localised States}

\section{Introduction}

In the past few years much attention has been drawn to the properties of shot noise in mesoscopic structures
\cite{Blanter!Buttiker}. So far experimental studies of shot noise have been primarily concentrated on
ballistic and diffusive systems \cite{QPC,OneThird}, with only a few exceptions
\cite{Schonenberger,KuznetsovHop} where shot noise in electron tunnelling or hopping between localised states
was investigated.

In short, mesoscopic barriers, where the condition $L/L_c<1$ is realised ($L_c$ is the correlation length
which represents the typical distance between the dominant hops in the hopping network \cite{SandE}), we have
studied shot noise in hopping. A suppression of shot noise in this regime was detected in
\cite{KuznetsovHop}: $S_I=F2eI$ where $F<1$ is the Fano factor. We have found that the result depends
significantly on the geometry of the sample: the suppression of shot noise for a wide sample (2D geometry) is
found to be much stronger than for a narrow (1D geometry) sample for the same sample length \cite{Roshko}. We
explain this by the reconstruction of the hopping network in 2D \cite{Raikh!Rusin,Savchenko!Raikh}: in a
short and wide sample the current is carried by a set of most conductive parallel hopping chains.

With further decreasing the sample length and temperature, resonant tunnelling through a single localised
state is seen. In this situation the suppression factor $F$ has been predicted to be $\left( \Gamma
_L^2+\Gamma _R^2 \right)/$ $\left( \Gamma _L+\Gamma _R\right) ^2$ \cite{Nazarov}, where $\Gamma _{L,R}$ are
the leak rates from the state to the left and right contacts. In our study of shot noise in resonant
tunnelling we have observed not only the suppression of noise, but also its significant enhancement
\cite{Safonov}. We have proved that this effect is caused by the Coulomb interaction between two parallel
resonant tunnelling channels, when two localised states carry the current in a correlated way. Experimental
observation of increased noise in this case is confirmed by theoretical calculations.

The experiment has been carried out on a \emph{n}-GaAs MESFET consisting of a GaAs layer of $0.15$ $\mu$m
thickness (donor concentration $N_d=10^{17}$ cm$^{-3}$). On the top of the GaAs layer Au gates are deposited
with dimensions $L=0.4$, $W=4$ $\mu $m, and $L=0.2$, $W=20$ $\mu $m, Fig.\ref{GVgT}a~(i). By applying a
negative gate voltage, $V_g$, a lateral potential barrier is formed between the ohmic contacts (source and
drain). Some gates contain a split of width $\omega =0.3$, 0.4 $\mu $m with the aim to define a
one-dimensional hopping channel.

\section{Suppression of shot noise in 1D and 2D hopping}\label{SecHopp}

\begin{figure}[t]
\centering \psfig{figure=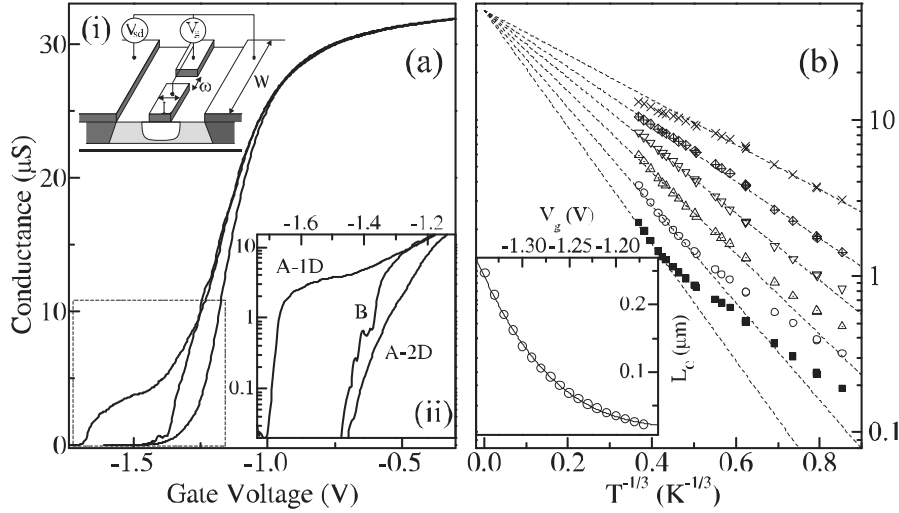,width=0.8\textwidth} \caption{(a) Conductance as a function of the gate
voltage for 2D and 1D configurations (samples A-2D and A-1D, B, respectively) at $T=4.2$ K. Insets: (i)
Cross-section of the transistor structure with two contacts and a split gate between them. (ii) Enlarged
$G\left( V_g\right) $ near the pinch-off shown by the dashed box on the main figure. (b) Temperature
dependence of the conductance of the sample A-2D for different gate voltages: $V_g=-1.18$ V (crosses), -1.22
V (diamonds), -1.25 V (down triangles), -1.28 V (up triangles), -1.31 V (circles), and -1.34 V (squares).
Solid lines are the fit by $T$-dependence of VRH. Inset: Correlation length for the hopping network as a
function of gate voltage.} \label{GVgT}
\end{figure}

We have studied the sample with two gates on the same structure ($L=0.4$ $\mu $m and $W=4$ $\mu $m) with
splits of different width: $\omega =0.3$ $\mu$m (sample A) and $\omega =0.4$ $\mu$m (sample B). The sample
with the narrow split shows different behaviour from one cooldown to another: two-dimensional (referred to as
A-2D) and one-dimensional (A-1D), Fig.\ref{GVgT}a. This difference in $G\left( V_g\right)$ for the 1D and 2D
configurations of the same sample can be caused by randomly trapped charge near the 1D constriction. When the
1D channel is blocked by strong fluctuations, electron transport is only possible under the continuous parts
of the gate and the sample is effectively 2D. In the 1D configuration electron transport through the split is
seen in Fig.\ref{GVgT}a (ii) as a characteristic bend in the conductance. The second sample (B) has a
$G\left( V_g\right) $ that does not look similar to either of the two curves of sample A.

The resistance of a macroscopic hopping network is determined by the most resistive (dominant) hops separated
by the distance of about $L_c$. To determine the characteristic length $L_c$ in the structure at different
gate voltages, the $T$-dependence of the conductance of sample A in its 2D configuration has been measured,
Fig.\ref{GVgT}b. In the gate voltage region from $-1.34$ to $-1.18$ V good agreement with the $T$-dependence
of variable range hopping (VRH) $G=G_0\exp\left( -\xi \right)$ is found at $T>4$ K (where $\xi =\left(
T_0/T\right)^{1/3}$, $T_0=\beta/\left( k_Ba^2g\right) $, $g$ is the density of states at the Fermi level,
$\beta =13.8$ and $a$ is the localisation radius). The parameter $\xi $ at $T=4.2$ K ranges from 2 to 6 as
the negative gate voltage is increased, reflecting the decrease of the density of states as the Fermi level
goes down. The correlation length $L_c\simeq a\xi ^{2.33}/2$ is presented in Fig.\ref{GVgT}b (inset).

In order to measure shot noise at $T=4.2$ K at a fixed $V_g$ different currents $I_{sd}$ are put through the
sample, and voltage fluctuations are measured by two low-noise amplifiers. The cross-correlated spectrum of
the two signals, $S_I \left( I_{sd}\right) $, is detected by a spectrum analyser -- this technique removes
uncorrelated noise of the amplifiers and leads. The shot noise power is determined from the flat part of the
excess noise spectrum above 30 kHz, where one can neglect the contribution of $1/f$ noise \cite{Roshko}.


In order to determine the Fano factor we used the following fit for excess noise: $S_I \left( I_{sd}\right)
=F2eI_{sd}$ $\coth \left( \frac{FeV_{sd}}{2k_B T }\right) -4k_BTG_S$, where $G_S$ is the ohmic conductance.
This expression describes the evolution of noise from thermal noise for $eV_{sd}\ll k_BT/F$ into shot noise
($eV_{sd}\gg k_BT/F$) with the Fano factor $F=1/N$, where $N$ is the number of the dominant barriers along
the chain \cite{derZiel,Korotkov1D}.

\begin{figure}[t]
\centering \psfig{figure=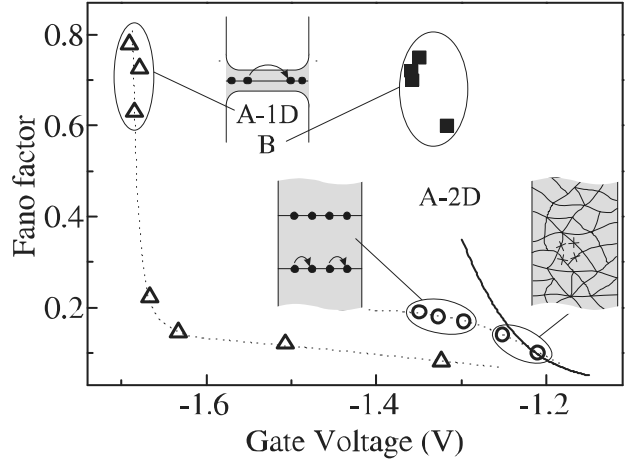,width=0.55\textwidth} \caption{Fano factor as a function of gate voltage
for different samples: A-2D (circles), A-1D (triangles) and B (squares), with a schematic representation of
hopping paths in each case. Small crosses show dominant hops. The solid line shows the $\left( L/L_c\right)
^{-1}$ dependence and dotted lines are guides to the eye.} \label{FanoVg}
\end{figure}

The Fano factor as a function of gate voltage for different structures is shown in Fig.\ref{FanoVg}. In the
2D case the Fano factor slowly increases from 0.1 to 0.2 with increasing negative gate voltage. According to
the $F=1/N$ model this change corresponds to a decrease in the number of dominant hops from $10$ to 5. The
value $L/L_c$ from Fig.\ref{GVgT}b (insert) is also plotted in Fig.\ref{FanoVg}, where in the range $V_g$
from -1.34 to -1.18 V we find agreement between the Fano factor and the number of the dominant hops in the
VRH network. With depleting the conducting channel the correlation length increases and approaches the length
of the sample, while the Fano factor shows a saturation around $F\sim 0.2$. The difference between $\left(
L/L_c \right) ^{-1}$ and $F$ confirms that electron transport at these gate voltages is dominated by chains
of most conductive hops \cite{Savchenko!Raikh,Raikh!Rusin}. High conductance of these `optimal' chains is
provided by the close position of localised states in them and the fact that all the barriers between the
states are equal.

In sample A-1D the 1D channel is only formed at $V_g< -1.3$ V, Fig.\ref{GVgT}a. In the range of $V_g$ from
-1.32 to -1.63 V, the Fano factor increases from 0.07 to 0.15 which corresponds to the number of dominant
hops $N\simeq 1/F=7$. With further increasing negative gate voltage ($V_g<-1.65$ V) the Fano factor in
Fig.\ref{FanoVg} rapidly increases to 0.8 as the distribution of the resistances of random hops in the 1D
channel is exponentially broad. In this case a single hop dominates the whole conductance of the 1D channel,
and the Fano factor is close to 1.

The conductance of the second sample (B) has not shown a clear definition of the 1D channel in
Fig.\ref{GVgT}a. For this sample shot noise in the range of $V_g$ from -1.36 V to -1.31 V has shown an
increase of the Fano factor from 0.6 to 0.8, Fig.\ref{FanoVg}. This large value compared with $F\sim 0.2$
expected for 2D hopping implies that, similar to A-1D, hopping in this sample occurs mainly through the 1D
split and is dominated by one or two hard hops. The differences in $G\left( V_g\right)$ of samples A-1D, A-2D
and B emphasise the importance of the random fluctuation potential in the formation of 1D channels
\cite{Davies}. If the random potential of impurities near the split is large, the conducting channel in the
split cannot be formed (as in the case of sample A-2D). Sample B has a larger width of the split, and in
several cooldowns the 1D channel has not been `blocked' by the random potential. In the hopping regime (at
$V_g<-1.31$ V) shot noise shows that the sample conduction is entirely determined by the split region.

It is interesting to note that for all studied samples, including the one with length 0.2 $\mu $m, the
Poisson value of the Fano factor, $F=1$, has never been observed. In Fig.\ref{FanoVg} the maximum Fano factor
fluctuates near $F\sim 0.7$. Although similar average values of $F$ have been predicted for electron
tunnelling through one impurity randomly positioned along the barrier ($F=0.75$ \cite{Nazarov}) and for
tunnelling through two impurities ($F\sim 0.71$ \cite{Korotkov2imp}), this fact deserves further attention.

\section{Enhancement of shot noise in resonant tunnelling via interacting localised states}

\begin{figure}[t]
\centering \psfig{figure=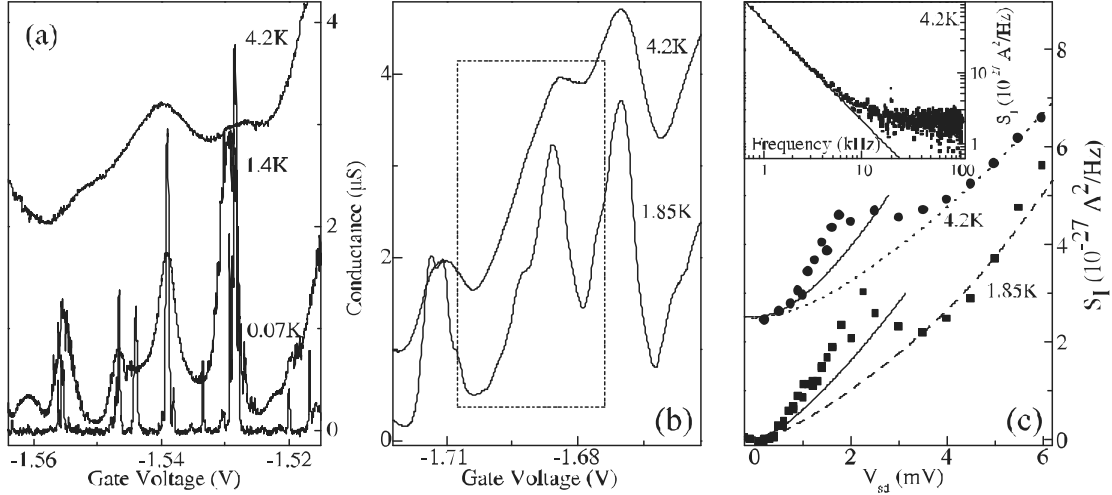,width=0.99\textwidth} \caption{(a) Typical resonant tunnelling peaks
in the ohmic conductance at different temperatures. (b) Conductance peaks in the region of $V_g$ where the
current noise has been measured. (c) Shot noise power as a function of $V_{sd}$: at $V_g=-1.6945$ V for
$T=1.85$ K and $V_g=-1.696$ V for $T=4.2$ K. Lines show the dependences $S_I\left( V_{sd}\right) $ expected
for resonant tunnelling through a single impurity, with $F=1$ (solid), $F=0.63$ (dashed), and $F=0.52$
(dotted). Inset: Excess noise spectrum at $V_g=-1.696$ V and $V_{sd}=1.5$ mV.} \label{Noise1}
\end{figure}

Shot noise in the case of resonant tunnelling through a single impurity has been studied on a 2D sample with
gate length 0.2 $\mu $m and width 20 $\mu $m. One can see in Fig.\ref{Noise1}a that as the temperature is
lowered the background conduction due to hopping decreases and the amplitude of the peaks increases, which is
a typical feature of resonant tunnelling through an impurity \cite{Fowler}. The box in Fig.\ref{Noise1}b
indicates the  range of $V_g$ where shot noise has been studied at $T=1.85$ K and 4.2 K. In Fig.\ref{Noise1}c
(inset) an example of the noise spectrum is shown at a gate voltage near the resonant tunnelling peak in
Fig.\ref{Noise1}b. The power spectral density of shot noise was determined at frequencies above 40 kHz where
the contribution of $1/f^\gamma $ noise (shown in Fig.\ref{Noise1}c (inset) by a solid line with $\gamma
=1.6$) can be totally neglected.

Fig.\ref{Noise1}c shows the dependence of the shot noise power on $V_{sd}$ at two temperatures. At small
biases ($V_{sd}<3$ mV) a pronounced peak in noise is observed, with an unexpectedly large Fano factor $F>1$.
At large biases ($V_{sd}>3$ mV), shot noise decreases to a conventional sub-Poisson value, $F\sim 0.6$. The
figure shows the dependence $S_I\left( V_{sd}\right) $ with different $F$ plotted using the phenomenological
expression for shot noise for resonant tunnelling through a single impurity (cf. Eq.(62) in
\cite{Blanter!Buttiker} and Eq.(11) in \cite{Nazarov}): $S_I=F2eI_{sd}\coth \left(
\frac{eV_{sd}}{2k_BT}\right) -F4k_BTG_S$. We have established that this increase of shot noise appears only
in a specific range of $V_g$ \cite{Safonov}. It will be shown below that the region $V_{sd}$-$V_g$ with
enhanced shot noise corresponds to the resonant current carried by two interacting impurities.

\begin{figure}[t]
\centering \psfig{figure=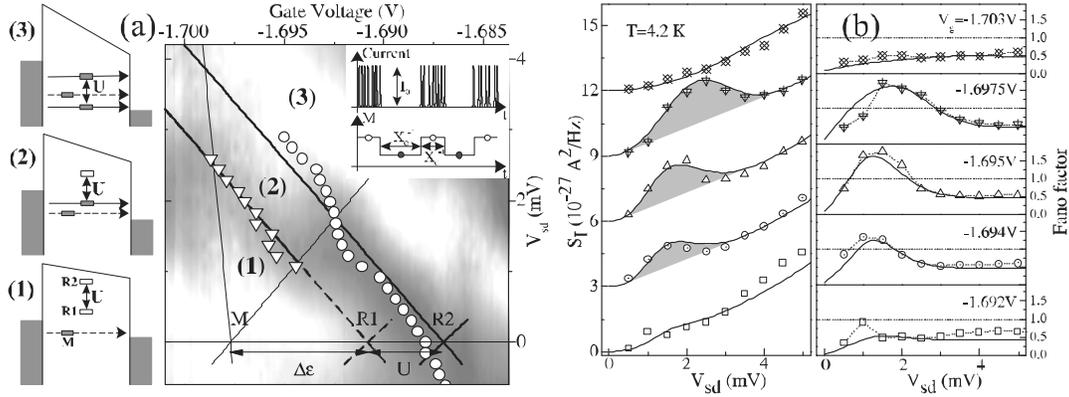,width=0.95\textwidth} \caption{(a) Left panel: Energy diagrams of the two
impurities for different positive $V_{sd}$: $V_{sd}^{(1)}<V_{sd}^{(2)}<V_{sd}^{(3)}$. Inset: Schematic
representation of the modulation of the current through impurity $R$ by changing the occupancy of modulator
$M$. Main part: Grey-scale plot of the differential conductance as a function of $V_{g}$ and $V_{sd}$ at
$T=1.85$ K (darker regions correspond to higher differential conductance, background hopping contribution is
subtracted). Lines show the positions of the conductance peaks of impurity $R$ and modulator $M$ obtained
from the fitting of the noise data in (b). (b) Shot noise and the corresponding Fano factor as a functions of
source-drain bias at different gate voltages. Solid lines show the results of the numerical calculation.}
\label{Grey}
\end{figure}

Consider two spatially close impurity levels, $R$ and $M$, separated in the energy scale by $\triangle
\epsilon$. If impurity $M$ gets charged, the level $R$ is shifted upwards by the Coulomb energy $U\sim
e^2/4\varepsilon _0\varepsilon r$, where $r$ is the separation between the impurities, Fig.\ref{Grey}a
(diagram 1). Thus, depending on the occupation of $M$ impurity $R$ can be in two states: $R1$ or $R2$. If
$V_{sd}$ is small enough, state $R2$ is above the Fermi level in the left contact and electrons are
transferred via $R1$ with the rates $\Gamma_{L,R}$ only when $M$ is empty, Fig.\ref{Grey}a (diagram 2).

It can be shown that for such correlated transport via two impurities, a significant increase of noise can be
seen if the leak rates of the two impurities are very different. For illustration, let us assume that,
independent of the occupancy of $R$, impurity $M$ (modulator) gets charged with rate $X_c$ and empties with
rate $X_e$. If $X_{e,c}\ll \Gamma _{L,R}$, the contribution of $M$ to the total current through the two
impurities is negligible. As a result, the current through impurity $R$ jumps randomly between two values:
zero and $I_0$, dependent on the occupancy of $M$, Fig.\ref{Grey}a (inset). If the bias is further increased,
the upper state $R2$ is shifted down and the modulation of the current via impurity $R$ vanishes,
Fig.\ref{Grey}a (diagram 3). In this modulation regime, the corresponding Fano factor can be written as
$F\simeq \frac{\Gamma _L^2+\Gamma _R^2}{\left(\Gamma _L+\Gamma _R\right)^2}+ 2\frac{\Gamma _L\Gamma
_R}{\Gamma _L+\Gamma _R}\frac{X_c}{\left(X_e+X_c\right)^2}$. The first term describes the conventional
(suppressed) Fano factor for one-impurity resonant tunnelling \cite{Nazarov}, whereas the second term gives
an enhancement of $F$. The latter comes from bunching of current pulses, which can be treated as an increase
of the effective charge in the electron transfers. Physically, this situation is similar to the increase of
flux-flow noise in superconductors when vortices move in bundles \cite{Clem}.

The generalisation of this simple model for any relation between $X$ and $\Gamma $ is based on the master
equation formalism \cite{Nazarov,Glazman!Matveev}. As a result of these calculations, the current through two
interacting impurities and the Fano factor are obtained as functions of the energy positions of the
impurities which are shifted with changing $V_{sd}$ and $V_g$.

We have shown that the increase of shot noise occurs exactly in the region of $V_g$-$V_{sd}$ where two
interacting impurities carry the current in a correlated way. Fig.\ref{Grey}a presents the grey scale of the
differential conductance versus $V_g$ and $V_{sd}$. When a source-drain bias is applied, a single resonant
impurity gives rise to two peaks in $dI/dV\left( V_g\right) $, which occur when the resonant level aligns
with the Fermi levels $\mu_{L,R}$. On the grey scale they lie on two lines crossing at $V_{sd}=0$. Therefore,
for impurity $M$ the central area between the thin lines of the cross at point $M$ in Fig.\ref{Grey}a
corresponds to its level being between $\mu_{L}$ and $\mu_{R}$, that is to the situation when its occupancy
changes with time. On the left of the central region the impurity is empty and on the right it is filled.

Experimentally, the cross of $M$ is not seen as the modulator gives a negligible contribution to the current
-- it is plotted in accordance with the analysis below. However, at small $V_{sd}$ a cross-like feature is
clearly seen near point $R2$ -- the exact positions of the maxima of the conductance peaks of this line are
indicated by circles. With increasing $V_{sd}$, a new parallel line $R1$ appears at $V_g \approx -1.694$ V
and $V_{sd} \approx 1$ mV, shifted to the left by $\triangle V_g \approx 4$ mV. This happens when the line
$R2$ enters the central area of cross $M$ -- the maxima of the conductance peaks of the new line are shown by
triangles. The lines $R1$ and $R2$ reflect the two states of impurity $R$ shifted due to the Coulomb
interaction with impurity $M$. The modulation of the current should then occur in region (2) of the central
area of the cross, Fig.\ref{Grey}a, which corresponds to diagram (2). In region (3) there is no modulation as
both states $R1$ and $R2$ can conduct, and in region (1) there is no current as the low state $R1$ is still
above $\mu _L$.

In Fig.\ref{Grey}b current noise and the Fano factor are presented as functions of $V_{sd}$ for different
$V_g$. It shows that indeed the increase of noise occurs only in region (2) in Fig.\ref{Grey}a. For a
quantitative analysis we have to take into account that in our experiment resonant tunnelling via state $R$
exists in parallel with the background hopping. Then the total Fano factor has to be expressed as $F=\left(
F_{RT}I_{RT}+F_{B}I_{B}\right) /\left( I_{RT}+I_{B}\right)$, where $F_{RT}$, $F_B$ and $I_{RT}$, $I_B$ are
the Fano factors and currents for resonant tunnelling and hopping, respectively. In order to get information
about the background hopping we have measured the current and noise at $V_g>-1.681$ V, i.e. away from the
resonant tunnelling peak under study in Fig.\ref{Grey}a.

The numerical results have been fitted to the experimental $dI/dV\left( V_{sd},V_g\right) $ and $S_I \left(
V_{sd}, V_g\right) $, Fig.\ref{Grey}a,b. The fitting parameters are the leak rates of $R$ and $M$ ($\hbar
\Gamma _{L}\simeq 394$ $\mu $eV, $\hbar \Gamma _{R}\simeq 9.8$ $\mu $eV, and $\hbar X_e\simeq 0.08$ $\mu $eV,
$\hbar X_c\simeq 0.16$ $\mu $eV ), the energy difference between $R$ and $M$ ($\triangle \varepsilon =1$
meV), and the Fano factor for the background hopping ($F_B=0.45$). This value of the Fano factor is expected
for shot noise in hopping through $N\sim 2-3$ potential barriers (1-2 impurities in series)
\cite{Roshko,Korotkov1D}. The coefficients in the linear relation between the energy levels $M$, $R$ and
$V_{sd}$, $V_g$ have also been found to match both the experimental data in Fig.\ref{Grey}b and the position
of lines $R1$ and $R2$ in Fig.\ref{Grey}a. The Coulomb shift ($U\sim 0.55$ meV) found from Fig.\ref{Grey}a
agrees with the estimation for the Coulomb interaction between two impurities not screened by the metallic
gate: $U\sim e^2/\kappa d \sim 1$ meV, where $d\sim 1000$ \AA \, is the distance between the gate and the
conducting channel.

It is interesting to note that the hopping background effectively hampers the manifestation of the enhanced
Fano factor $F_{RT}$: the largest experimental value of $F$ in Fig.\ref{Grey}b (at $V_g=-1.6975$ V) is
approximately $1.5$, while a numerical value for RT at this $V_g$ is $F_{RT} \approx 8$.

\begin{figure}[t]
\centering \psfig{figure=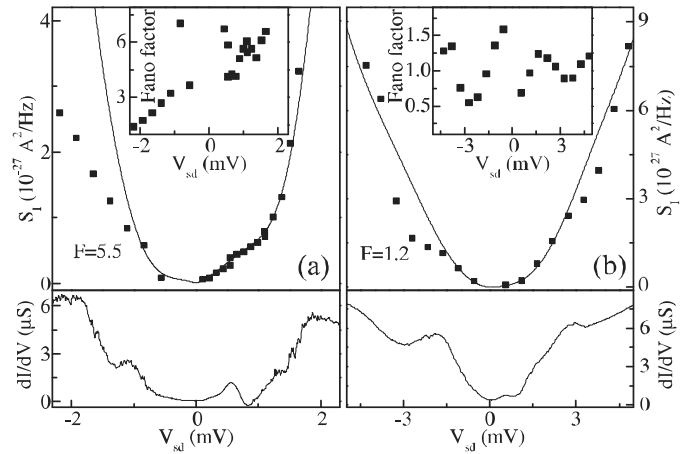,width=0.6\textwidth} \caption{Shot noise power and differential conductance
as a function of the current for the 2D sample with length 0.2 $\mu $m at different temperatures: (a)
$T=0.07$ K and (b) $T=1.4$ K. Solid lines show a fit for shot noise for resonant tunnelling through a single
impurity.} \label{RT}
\end{figure}

With further decreasing the temperature down to $T=70$ mK we have seen a significant increase of the Fano
factor up to $F\sim 5$, Fig.\ref{RT}. This regime corresponds to the quantum case $\hbar \Gamma\gg k_BT$
where the master equation approach is not applicable. We can assume that this increase is also related to the
interaction between different resonant tunnelling channels, although more theoretical input is required to
understand this effect. There is another interesting feature in Fig.\ref{RT}: the increase of shot noise
slows down at the position of the conductance peak, so that the Fano factor shows a decrease at the points,
where the new current channel is switched on, Fig.\ref{RT} (insets).

\section{Conclusion} \label{sect:latex}

The details of hopping transport in 1D and 2D have been investigated by measuring shot noise. It is shown
that suppression of shot noise contains information about the geometry of hopping paths and the distribution
of hopping barriers in them. In one-impurity resonant tunnelling we have observed a significant increase of
shot noise. It it shown that this increase is caused by the Coulomb interaction between two close resonant
tunnelling channels. In general, we have demonstrated that shot noise is a valuable tool for investigations
of electron transport in the insulating regime of conduction.

\end{document}